\documentclass{aastex63}

\usepackage{amsmath}
\usepackage{amssymb}
\usepackage{graphicx}
\begin{document}

\title{Annihilation of Magnetic Islands at the Top of Solar Flare Loops}

\correspondingauthor{Xin Cheng}
\email{xincheng@nju.edu.cn}

\author[0000-0001-9863-5917]{Yulei Wang}
\affiliation{School of Astronomy and Space Science, Nanjing University, Nanjing 210023, People's Republic of China}
\affiliation{Key Laboratory for Modern Astronomy and Astrophysics (Nanjing University), Ministry of Education, Nanjing 210023, People's Republic of China}

\author[0000-0003-2837-7136]{Xin Cheng}
\affiliation{School of Astronomy and Space Science, Nanjing University, Nanjing 210023, People's Republic of China}
\affiliation{Key Laboratory for Modern Astronomy and Astrophysics (Nanjing University), Ministry of Education, Nanjing 210023, People's Republic of China}

\author[0000-0002-4978-4972]{Mingde Ding}
\affiliation{School of Astronomy and Space Science, Nanjing University, Nanjing 210023, People's Republic of China}
\affiliation{Key Laboratory for Modern Astronomy and Astrophysics (Nanjing University), Ministry of Education, Nanjing 210023, People's Republic of China}

\author[0000-0003-3041-2682]{Quanming Lu}
\affiliation{CAS Key Laboratory of Geospace Environment, Department of Geophysics and Planetary Science, University of Science and Technology of China, Hefei 230026, People's Republic of China}
\affiliation{CAS Center for Excellence in Comparative Planetology, People's Republic of China}

\begin{abstract}
The dynamics of magnetic reconnection in the solar current sheet (CS) is studied by high-resolution 2.5-dimensional MHD simulation.
With the commence of magnetic reconnection, a number of magnetic islands are formed intermittently and move quickly upward and downward along the CS.
When colliding with the semi-closed flux of flare loops, the downflow islands cause a second reconnection with a rate even comparable with that in the main CS.
Though the time-integrated magnetic energy release is still dominated by the reconnection in main CS, the second reconnection can release substantial magnetic energy, annihilating the main islands and generating secondary islands with various scales at the flare loop top.
The distribution function of the flux of the second islands is found to follow a power-law varying from $f\left(\psi\right)\sim\psi^{-1}$ (small scale) to $\psi^{-2}$ (large scale), which seems to be independent with background plasma $\beta$ and if including thermal conduction.
However, the spatial scale and the strength of the termination shocks driven by main reconnection outflows or islands decrease if $\beta$ increases or thermal conduction is included.
We suggest that the annihilation of magnetic islands at the flare loop top, which is not included in the standard flare model, plays a non-negligible role in releasing magnetic energy to heat flare plasma and accelerate particles.
\end{abstract}


\section{Introduction} \label{sec:intro}

Magnetic reconnection, a fundamental energy-releasing process in magnetized plasma, is believed to be the core mechanism
driving solar eruptions including solar flares, coronal mass
ejections (CMEs), and other solar activities. According to the standard flare model, the sheared magnetic field is stretched and forms a large-scale thin current
sheet (CS) in the wake of the eruption \citep{KoppRA_1976_727,LinJun_2000_709,LinJun_2015_723}.
The reconnection in the CS in turn drives the CME eruption and the formation of the flare loops.
Theoretically, the squeezed thin CS is unstable as the tearing mode instability grows \citep{PriestEric_2000_633,PucciFulvia_2013_896,LoureiroNF_2015_725}.
Especially, in the corona of high Lundquist-number, the plasmoid instability will dominate the dynamics of magnetic
reconnection \citep{SamtaneyR_2009_800,BhattacharjeeA_2009_682,HuangYiMin_2010_634,HuangYiMin_2011_636,HuangYiMin_2013_623,HuangYiMin_2012_635,HuangYiMin_2017_611,LoureiroNF_2015_725,ShenChengCai_2011_619,ZhaoXiaozhou_2020_679}.
The classical CS predicted by Sweet-Parker is
thus split and the reconnection enters into
the fast scheme and generates a number of magnetic islands.

Magnetic islands are believed to be closely related to fine structures in the CS \citep{ShenChengCai_2011_619,LinJun_2015_723}.
The coalescence of magnetic islands can further produce secondary CSs and islands \citep{BartaMiroslav_2011_822}.
Various researchers have analyzed the cascading law of magnetic island flux theoretically or numerically, which predicts that
the distribution function of magnetic island flux $f\left(\psi\right)$ in the CS follows a power-law between $\psi^{-1}$ to $\psi^{-2}$ \citep[e.g.,][]{UzdenskyDA_2010_873,HuangYiMin_2012_635,ShenChengcai_2013_867,LynchBJ_2016_864,YeJing_2019_706}. The fragmented and turbulent characteristics of the CS have also been suggested by recent observations \citep[e.g.,][]{ChengXin_2018_341}.

The flare loop top, interacting with the reconnection downflows, also shows complex dynamics.
The turbulent characteristics of the loop top have been studied by high-resolution simulations \citep[e.g.,][]{CaiQiangwei_2019_872,YeJing_2020_653,CaiQiangwei_2021_852}.
The generation of flare quasi-periodic pulsations (QPPs) and the supra-arcade downflows (SADs) above the flare loop top is interpreted to be closely related to the fast reconnection downflows \citep{TakasaoShinsuke_2016_716,GuoLJ_2014_609}.
Abundant MHD shocks have been predicted by simulations \citep{TakasaoShinsuke_2015_717}.
Especially, as an important prediction of the standard solar flare model, the fast mode termination shocks (TSs), formed when the downflows hit the relatively steady high-density structure at the loop top, are believed to be an effective mechanism to accelerate particles \citep{TsunetaSaku_1998_791,ShenChengcai_2018_705,KongXiangLiang_2019_617,KongXiangLiang_2020_719,RuanWenzhi_2020_737}, which are supported by radio imaging observations \citep{AurassHenry_2004_730,ChenBin_2015_641}.
The downflow magnetic islands also affect the dynamics of the flare loops \citep{YeJing_2020_653,CaiQiangwei_2019_872}, the formation of TSs \citep{ShenChengcai_2018_705}, and even the acceleration of energetic electrons \citep{KongXiangLiang_2020_719}.

Nevertheless, mainly owing to the limitation of observations and simulations, magnetic reconnection in the CS and its relation to the dynamics at the flare loop top are still far from being fully understood.
In this paper, we perform high-resolution 2.5-dimensional MHD simulations of the CS reconnection in a high-Lundquist-number and low-$\beta$ coronal environment and focus on the dynamics of magnetic islands evolving towards flare loops.
We find that the downflow magnetic islands quickly annihilate after colliding with the tip of cusp-shaped flare loops.
Such a second reconnection process is characterized by horizontal CSs, smaller-scale islands, and a comparable reconnection rate with that in the main CS.
In Section \ref{sec:method}, we describe our method. Section \ref{sec:results} presents the main results, which are followed by a summary and discussion.

\section{Method\label{sec:method}}

In this work, the MHD equation we solve is as follows:
\begin{eqnarray}
\frac{\partial\rho}{\partial t}+\nabla\cdot\left(\rho{\bf u}\right) & = & 0\,,\nonumber \\
\frac{\partial\left(\rho{\bf u}\right)}{\partial t}+\nabla\cdot\left(\rho{\bf u}{\bf u}-{\bf BB}+P^{*}\right) & = & 0\,,\nonumber \\
\frac{\partial e}{\partial t}+\nabla\cdot\left[\left(e+P^{*}\right){\bf u}-{\bf B}\left(\bf{B}\cdot{\bf u}\right)\right] & = & \nabla\cdot\left(\kappa_{\parallel}\hat{\bf{b}}\hat{\bf{b}}\cdot\nabla T\right)\,,\label{eq:MHD}\\
\frac{\partial{\bf B}}{\partial t}-\nabla\times\left(\bf{u}\times\bf{B}\right) & = & -\nabla\times\left(\eta{\bf J}\right)\,,\nonumber \\
{\bf J} & = & \nabla\times{\bf B}\,,\nonumber
\end{eqnarray}
where, $P^*=p+B^2/2$, $e=p/\left(\gamma-1\right)+\rho u^2/2+B^2/2$, $\kappa_\parallel=\kappa_0T^{2.5}$, and standard notations of variables are used. In our model, only the thermal conduction (TC) along magnetic field is considered which is much more important than perpendicular components and $\kappa_0$ is set as $10^6\,\mathrm{erg\cdot s^{-1}\,cm^{-1}\,K^{-3.5}}$ \citep{YokoyamaTakaaki_2001_731}.
Here, all variables have been normalized according to constant units.
The unit of space is chosen as the scale of simulation region $L_0=5\times 10^9\,\mathrm{cm}$ that is comparable with the value for a typical observed flare event \citep[see,][]{ChengXin_2018_341}.
We also assume that the coronal plasma is composed of fully ionized hydrogen with the electron density $n_e=10^{10}\,\mathrm{cm^{-3}}$ and the averaged particle mass $\bar{m}=0.5m_p=8.36\times 10^{-25}\,\mathrm{g}$, which gives the unit of mass density $\rho_{0}=2n_e\bar{m}=1.67\times10^{-14}\,\mathrm{g/cm^{3}}$ \citep{PriestEric_2000_633}. Here, $m_p$ is the mass of proton.
The unit of magnetic strength is set as a typical coronal value, namely, $B_{0}=20\,\mathrm{Gauss}$ \citep[also see][]{ChenPengFei_1999_689,YeJing_2020_653}.
Based on $L_0$, $\bar{m}$, $\rho_0$, and $B_0$, normalized units of other variables are deduced as $u_{0}=B_{0}/\sqrt{\mu_{0}\rho_{0}}=4.36\times 10^7\,\mathrm{cm/s}$, $t_{0}=L_0/u_0=114.61\,\mathrm{s}$, $p_{0}=\rho_0u_0^2=3.18\,\mathrm{Pa}$, $T_0=\bar{m}u_0^2/k_B=11.52\,\mathrm{MK}$, $\kappa_{\parallel 0}=k_B\rho_0 u_0L_0/\bar{m}=6.02\times10^{11}\,\mathrm{erg/s\cdot cm\cdot K}$, and $\eta_{0}=L_{0}u_{0}=2.18\times 10^{17}\,\mathrm{cm^2/s}$, where $\mu_{0}$ is the magnetic permeability in vacuum and $k_B$ denotes the Boltzmann constant.
The ratio of specific heat is $\gamma=5/3$.
Under this configuration, the unit time, velocity, and temperature are also on the same order as the observational ones given by \cite{ChengXin_2018_341}.

The initial magnetic field is set according to the CSHKP model \citep{ChenPengFei_1999_689,YeJing_2020_653}
\begin{eqnarray}
B_{x} & = & 0\,,\nonumber \\
B_{y} & = & \begin{cases}
\sin\left(\frac{\pi x}{2\lambda}\right)\,, & \left|x\right|\leqslant\lambda\,,\\
1\,, & x>\lambda\,,\\
-1\,, & x<-\lambda\,,
\end{cases}\label{eq:mag}\\
B_{z} & = & \sqrt{1-B_{x}^{2}-B_{y}^{2}}\,,\nonumber
\end{eqnarray}
where $\lambda$ denotes the half-width of the CS, which is set as 0.1.
To approximate the chromosphere, transition region, and corona, the initial density distribution follows \cite{TakasaoShinsuke_2015_717}, namely,
\begin{equation}
    \rho\left(y\right)=\rho_{chr}+\frac{\rho_{cor}-\rho_{chr}}{2}\left[\mathrm{tanh}\left(\frac{y-h_{chr}}{l_{tr}}+1\right)\right]\,,\label{eq:rho}
\end{equation}
where, $\rho_{cor}=1$, $\rho_{chr}=10^5$, $h_{chr}=0.1$, and $l_{tr}=0.02$.
Fast reconnection is initially triggered by a localized anomalous resistivity and thus the resistivity distribution is
\begin{equation}
\eta=\begin{cases}
\eta_{b}+\eta_{a}\mathrm{exp}\left[-\frac{x^{2}+\left(y-h_{\eta}\right)^{2}}{l_{\eta}^{2}}\right]\,, & t\leq t_{\eta}\,,\\
\eta_{b}\,, & t>t_{\eta}\,,
\end{cases}\label{eq:eta}
\end{equation}
where, $\eta_a=5\times10^{-4}$, $h_\eta=0.5$, $l_\eta=0.03$, $t_\eta=5$, and the background resistivity is uniformly set as $\eta_b=5\times10^{-6}$ to simulate a high-Lundquist-number environment.
Physically, the localized anomalous resistivity can be caused by microscopic instabilities in the current sheet such as the lower hybrid drift and/or ion acoustic instabilities which can boost the process of reconnection \citep{YokoyamaTakaaki_2001_731}. Because $\eta_a\gg \eta_b$, the anomalous resistivity dominates the evolution during $t<t_\eta$, which provides similar initial states at $t=t_\eta$ for the cases listed in Tab.\,\ref{tab:Cases}.
The background pressure $p_b$ is uniform to obtain a static initial state and is assigned as shown in Tab.\,\ref{tab:Cases}.
The boundary conditions are arranged as follows.
The left ($x=-1$) and right ($x=1$) are free boundaries, the top ($y=4$) is no-inflow boundary, and the bottom ($y=0$) is symmetric boundary.
To reduce the influences of numerical boundaries, we only use data in the region defined by $x\in\left[-0.5,0.5\right]$ and $y\in\left[0,2\right]$ for analysis.

\begin{table}[h]
    \centering
    \begin{tabular}{cccc}
    \hline
    \textbf{Cases} & $p_b$ & $\beta$ & \textbf{TC}\tabularnewline
    \hline
    1 & 0.02 & 0.04 & No\tabularnewline
    2 & 0.05 & 0.10 & No\tabularnewline
    3 & 0.10 & 0.20 & No\tabularnewline
    4 & 0.02 & 0.04 & Yes\tabularnewline
    \hline
    \end{tabular}
    \caption{Configurations of all simulation cases.\label{tab:Cases}}
\end{table}

The above system is numerically solved with Athena 4.2 \citep{StoneJamesM_2008_833}.
We use the HLLD Riemann solver \citep{MiyoshiTakahiro_2005_876}, 3-order piecewise parabolic flux reconstruction algorithm \citep{StoneJamesM_2008_833}, and the Corner Transport Upwind (CTU) method \citep{GardinerThomasA_2008_877} to solve the conservation part of Eq.\,\ref{eq:MHD}. The resistivity and TC are calculated by the explicit operator splitting method.
To suppress numerical dissipation and get a uniformly high resolution \citep{ShenChengCai_2011_619,YeJing_2020_653},
we set high-precision uniform Cartesian grids, namely, 3840 and 7680 grids on $x$ and $y$ directions, respectively. The corresponding pixel scales in two directions are $\Delta x=\Delta y=26\,\mathrm{km}$.
The maximum simulation time in our simulation is $t_{max}=15$, which
corresponds to $28.65$ minutes in physical time.

\section{Results\label{sec:results}}

\subsection{Overview\label{subsec:overview}}

\begin{figure}
\begin{centering}
\includegraphics[scale=0.8]{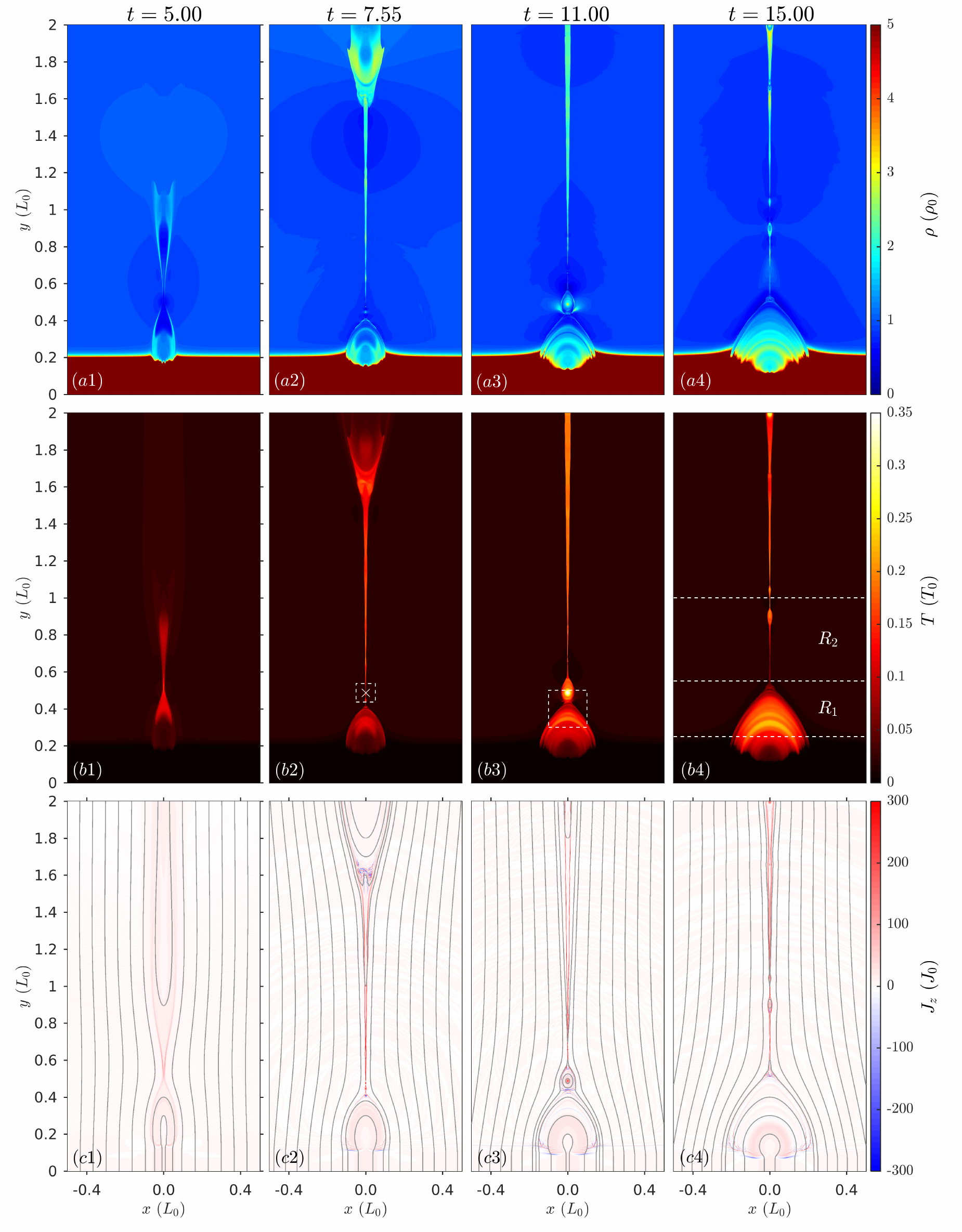}
\par\end{centering}
\caption{Snapshots of mass density $\rho$, temperature $T$, and
current density $J_{z}$ at four typical moments for Case 4. The 1st column at
$t=5$ shows the initial status when the localized anomalous resistivity is shut down.
The erupting plasmoid, the stretched main CS, and the flare loop ``seed'' are clearly displayed by the 2nd column at $t=7.55$.
The 3rd column records the moment when a relatively large magnetic island starts to collide with the loop-top flux at $t=11$.
The 4th column exhibits the final state of the flare loop systems at $t=15$.
The x-marker and the dashed box in (b2) denote respectively the principal X-point in the main CS and the inflow region for calculating the averaged magnetic strength and Alfv\'{e}n speed in Eq.\,\ref{eq:R}.
The dashed box in (b3) marks the loop-top region analyzed in Fig.\,\ref{fig3}.
The three dashed horizontal lines in (b4), namely, $y=0.25$, $y=y_{lt}\left(t\right)$, and $y=1$, define the loop-top region, $R_1$, and the principal reconnection site in the main CS, $R_2$. Here, $y_{lt}\left(t\right)$ denotes the height of the top of the cusp-shaped loops, which dynamically changes with the evolution of the system.\label{fig1}}
\end{figure}

Figure \ref{fig1} shows four typical snapshots of density $\rho$, temperature $T$, and out-of-plane current $J_{z}$ for Case 4.
Before $t=5$, the reconnection is dominated by the anomalous resistivity at $y=0.5$ where the main CS is squeezed.
After $t=5$, the anomalous resistivity is set to zero and the reconnection is determined by the uniform background resistivity $\eta_b$.
At $t=5$, a plasmoid grows above the X-point and starts to move upward, while the lower part of the main CS forms a "seed" of flare loops with two foot-points line-tied in the high-density chromosphere (1st column of Fig.\,\ref{fig1}).
As the ejection of the upper plasmoid, the main CS is rapidly stretched (2nd column of Fig.\,\ref{fig1}).
When its aspect ratio is large enough, the main CS is quickly split by the plasmoid instability \citep[also see][]{ShenChengCai_2011_619}.
The downflow magnetic islands carrying substantial kinetic and thermal energies are intermittently formed and then collide with the cusp-shaped flare loops.
At $t=11$, a relatively large magnetic island starts to interact with the loop-top flux (3rd column of Fig.\,\ref{fig1}) and disappears at $t=11.65$.
Fed by the downflow islands, the flare loops also manifest an obvious expansion.
At $t=15$, the simulation stops and the flare loops evolve into their final states (4th column of Fig.\,\ref{fig1}).
The distributions of density and temperature are consistent with the results of \citet{YeJing_2020_653} and highly resemble observed flare loops \citep[e.g.,][]{SunJQ_2014_878}.

\subsection{Reconnection in the Main CS\label{subsec:csmr}}

\begin{figure}
\begin{centering}
\includegraphics[scale=0.8]{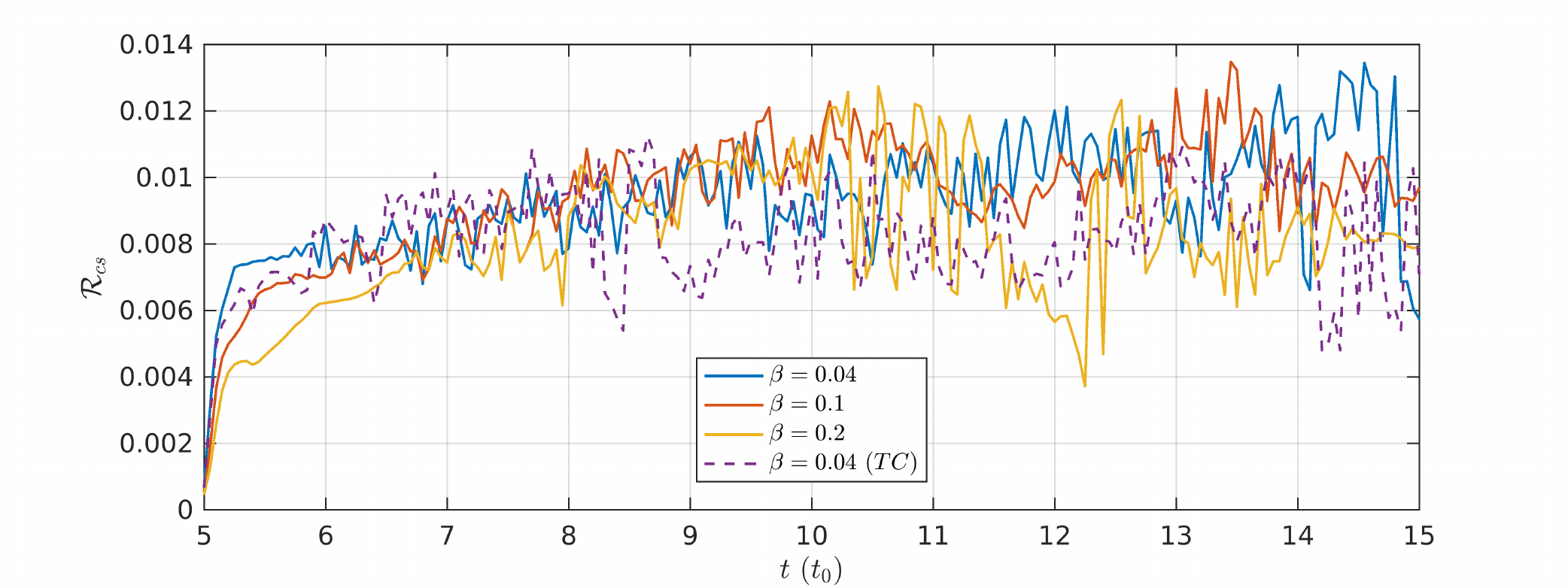}
\par\end{centering}
\caption{Evolutions of the magnetic reconnection rate in the main CS. The blue, orange, and yellow solid curves show $\mathcal{R}_{cs}\left(t\right)$ for Cases 1, 2, and 3, respectively, while the purple dashed curve presents Case 4 in which TC is considered. Here we only exhibit the reconnection rate after $t=5$, because it is dominated by the anomalous resistivity before $t=5$.\label{fig2}}
\end{figure}

We first estimate the reconnection rate in the main CS as shown by the area $R_2$ in Fig.\,\ref{fig1}(b4)).
It is represented by the reconnection electric field at the principal X-point \citep[see also,][]{YokoyamaTakaaki_2001_731}, namely,
\begin{equation}
    \mathcal{R}\left(t\right)=\frac{\max\left(\eta_b\left|J_{zxp}\left(t\right)\right|\right)}{B_{in}\left(t\right)u_{in}\left(t\right)}\,,\label{eq:R}
\end{equation}
where, $J_{zxp}$ denotes the out-of-plane current density at X-points, $\max\left(\cdot\right)$ means taking the maximum value in the target CS region, and
$\mathcal{R}$ is normalized by the product of inflow magnetic strength $B_{in}$ and Alfv\'{e}n speed $u_{in}$ which are averaged in the adjacent region of the principal X-point defined by $x\in\left[-0.05,0.05\right]$ and $y\in\left[y_{xp}-0.05,y_{xp}+0.05\right]$ (see, Fig.\,\ref{fig1}(b2)). Here, $y_{xp}$ denotes the y-coordinate of the principal X-point.
The null points are determined by the method developed by \cite{ParnellCE_1996_860}.

For each simulation case, the reconnection rate in the main CS, $\mathcal{R}_{cs}$, shows a similar evolution trend (Fig.\,\ref{fig2}). To be specific, the average value of $\mathcal{R}_{cs}$ first rises quickly and then reaches its peak value which is on the order of $0.01$.
We can see significant oscillations in $\mathcal{R}_{cs}$ curves which correspond to the formation of magnetic islands \citep{YokoyamaTakaaki_2001_731}.
The overall evolution of the reconnection rate in the main CS is consistent with previous numerical results of 2D reconnection \citep[e.g.,][]{NiLei_2012_768,HuangYiMin_2016_608,ZenitaniSeiji_2020_631}.
Differing $\beta$ mainly affects the initial rising stage before $t=7$.
With the increase of $\beta$, the reconnection rate at this stage becomes smaller.
However, when the reconnection is fully developed ($t>10$), the evolution of $\mathcal{R}_{cs}$ is less affected by $\beta$.
After including TC, the reconnection rate before $t=6$ gets smaller and the average value also decreases slightly after $t=10$ (Fig.\,\ref{fig2}).

\subsection{Second Reconnection of Magnetic Islands\label{subsec:ltmr}}

\begin{figure}
\begin{centering}
\includegraphics[scale=0.45]{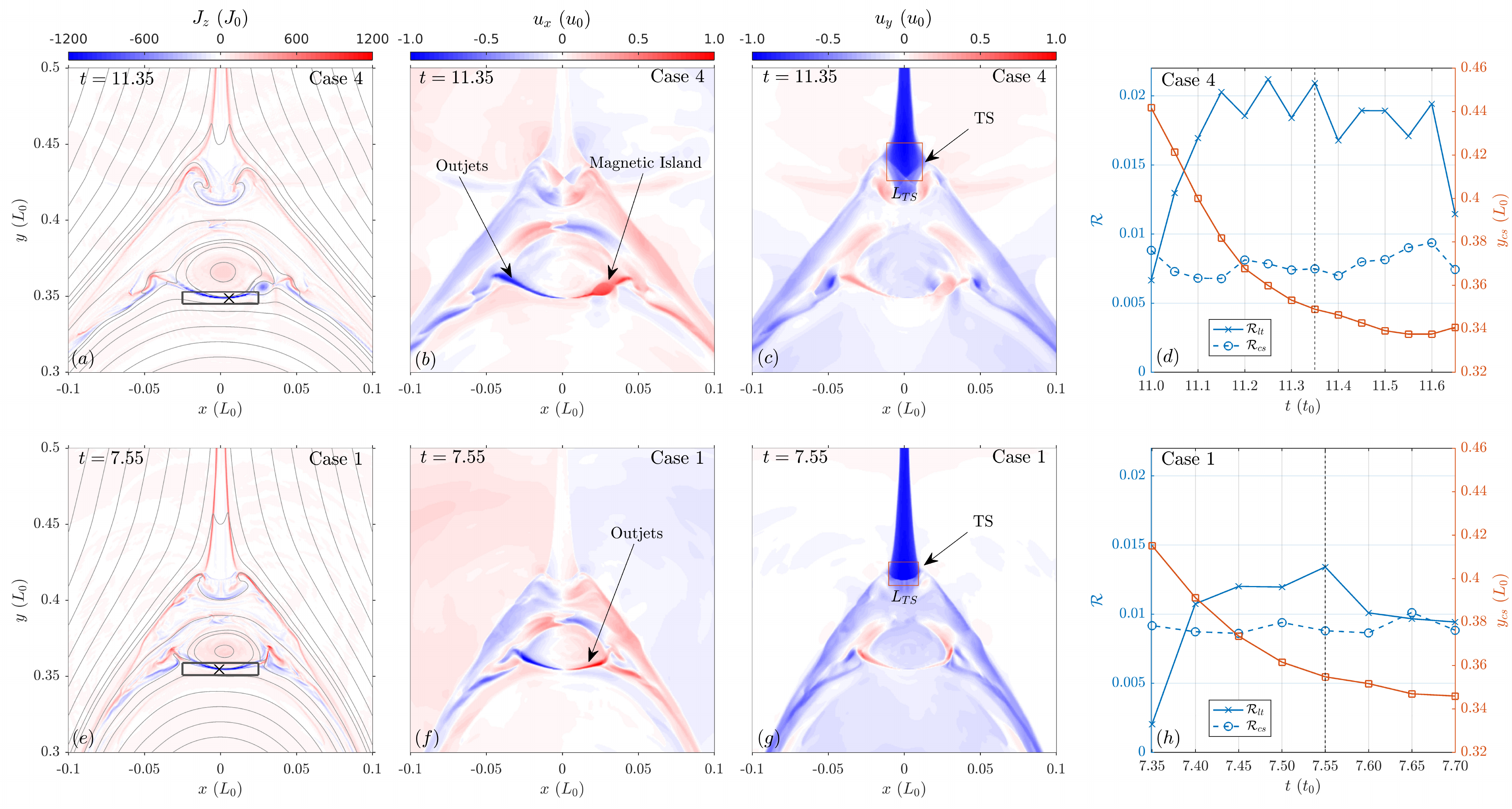}
\par\end{centering}
\caption{Two typical magnetic island annihilation events in Case 4 and Case 1.
Panels (a), (b), and (c) draw the current, the horizontal velocity, and the vertical velocity around the peak time of $\mathcal{R}_{lt}$.
In panel (d), the blue and orange solid curves plot the reconnection rate $\mathcal{R}_{lt}$ and the height of CSs, respectively.
And the dashed blue curve depicts the reconnection rate $\mathcal{R}_{cs}$ in the main CS.
The box and "x" symbol in panels (a) mark the horizontal CS and the principal X-point, respectively.
The principal TS is marked by the box in panel (c) and the spatial scale of TS is approximately represented by its scale along the $x$-direction. Panels (e)-(f) display a similar event in Case 1 where TC is not included. \label{fig3}}
\end{figure}

To study the loop-top annihilation of magnetic islands in detail, we analyze a typical event that appears at $t\in\left[11,11.65\right]$ in Case 4 (3rd column of Fig.\,\ref{fig1}).
As the magnetic island collides with the loop-top flux, a horizontal current sheet forms (see, Fig.\,\ref{fig1}(c3) and Fig.\,\ref{fig3}(a)).
Correspondingly, a pair of horizontal jets appear and quickly move along two opposite directions (Fig.\,\ref{fig3}(b)).
The peak speed of these jets reaches $0.8$, which is of a similar order as the fastest downflow in the main CS.
When these horizontal jets encounter the edge of the flare loop, they start to move downward, leading to the increase of downward speed $u_{y}$ (Fig.\,\ref{fig3}(c)), as observed by \citet{TianHui_2014_753}.
It is thus clearly revealed that the second reconnection enables the magnetic island annihilation at the flare loop top.
During the second reconnection of islands, new small-scale magnetic islands are also generated in the loop-top CSs (see, Fig.\,\ref{fig3}(a) and (b)), similar to the ``fragmenting coalescence'' picture proposed by \cite{BartaMiroslav_2011_822}.

To estimate the reconnection rate of the horizontal CS, we trace it using a box with the height of $0.008$ and width of $0.05$ (see, Fig.\,\ref{fig3}(a)).
After locating all X-points in this region, we use Eq.\,\ref{eq:R} to estimate $\mathcal{R}_{lt}$, the reconnection rate in the horizontal CS.
It is shown that, as the magnetic island annihilates, the height of the horizontal CS keeps decreasing (Fig.\,\ref{fig3}(d)).
While for the reconnection rate within, it first increases, then reaches its peak value, and finally decreases (Fig.\,\ref{fig3}(d)).
Obviously, the peak reconnection rate is comparable with that in the main vertical CS.
We also analyze a similar event in Case 1 where TC is not considered (see, Fig.\,\ref{fig3}(e)-(h)).
It is found that the evolution of horizontal CS, outjets and the reconnection rate are similar to Case 4.

\begin{figure}
\begin{centering}
\includegraphics[scale=0.8]{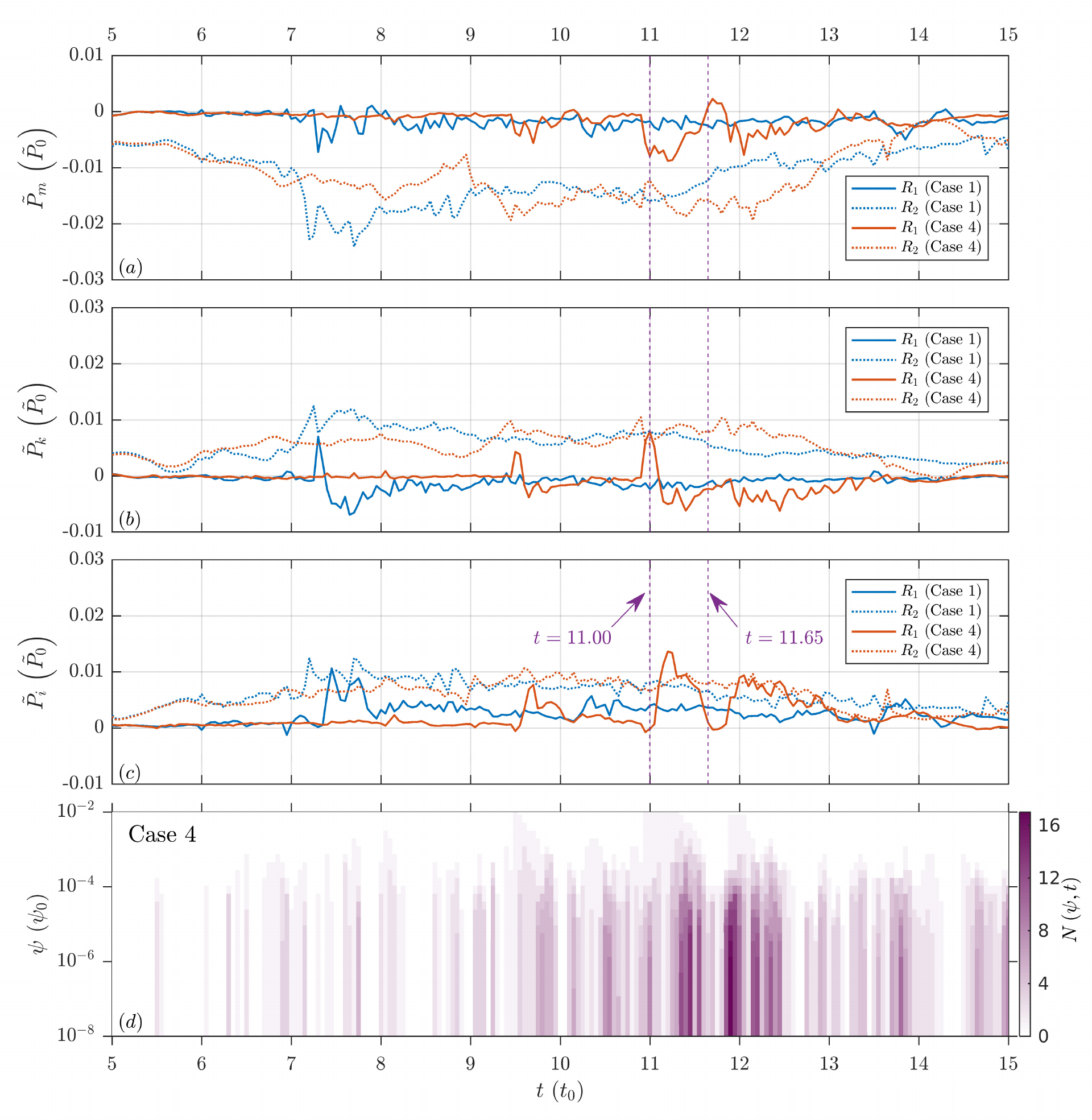}
\par\end{centering}
\caption{
Illustrations of the energy conversion and magnetic island distribution of the second reconnection at the flare loop top.
Panels (a), (b), and (c) show respectively the evolutions of the effective changing rate of magnetic, kinetic, and internal energy in the main CS and at the flare loop top.
The orange (blue) curves show the results with (without) TC.
For each case, the solid and dashed curves depict the evolution of $\tilde{P}$ in regions $R_1$ and $R_2$, respectively.
The physical unit of $\tilde{P}$ is $6.94\times10^{18}\,\mathrm{erg/\left(cm\cdot s\right)}$.
Panel (d) depicts the CDF of magnetic island flux $N\left(\psi,t\right)$ at loop-top region $R_1$ in Case 4.
To obtain $N\left(\psi,t\right)$, we set $30$ bins spaced uniformly in $\log\left(\psi\right)$ over the range $\psi\in\left[10^{-8},0.0215\right]$ in units of $\psi_0=10^{11}\,\mathrm{Mx/cm}$.\label{fig4}
}
\end{figure}

We further study the energy conversion during the second reconnection at the loop-top region and compare it with that in the main CS.
In an area $S$, the effective changing rates of magnetic, kinetic, and internal energies are respectively defined by \citep{ForbesTG_1988_816}
\begin{eqnarray}
\tilde{P}_m & = & \frac{\mathrm{d}}{\mathrm{d}t}\int_{S}\frac{B^{2}}{2}\mathrm{d}x\mathrm{d}y+\oint_{\partial S}\left({\bf E}\times{\bf B}\right)\cdot{\bf n}\mathrm{d}l\,,\label{eq:Pm} \\
\tilde{P}_k & = & \frac{\mathrm{d}}{\mathrm{d}t}\int_{S}\frac{\rho u^{2}}{2}\mathrm{d}x\mathrm{d}y+\oint_{\partial S}\frac{\rho u^2}{2}{\bf u}\cdot{\bf n}\mathrm{d}l\,,\label{eq:Pk}\\
\tilde{P}_i & = & \frac{\mathrm{d}}{\mathrm{d}t}\int_{S}\frac{p}{\gamma-1}\mathrm{d}x\mathrm{d}y+\oint_{\partial S}\frac{\gamma p}{\gamma-1}{\bf u}\cdot{\bf n}\mathrm{d}l\,,\label{eq:Pi}
\end{eqnarray}
where, $\partial S$ denotes the boundary, ${\bf E}=\eta{\bf J}-{\bf u}\times{\bf B}$ is the electric field, ${\bf n}$ is the unit normal vector on the boundary, and $dl$ denotes the line element.
The negative value of $\tilde{P}_m$ implies the release of magnetic energy.
We calculate the evolutions of $\tilde{P}_m$, $\tilde{P}_k$, and $\tilde{P}_i$ at the loop-top region, $R_1$, and that in the main CS, $R_2$ (see, Fig.\,\ref{fig1}(b4)).
It's found that, during most of the time, the magnetic energy release is dominated by the main CS reconnection, especially during the early stage (Fig.\,\ref{fig4}(a)).
However, the second reconnection at the loop-top region is able to further release magnetic energy (see, negative spikes of solid curves in Fig.\,\ref{fig4}(a)).
For example, in Case 4, during $t\in\left[11,11.65\right]$, the effective magnetic energy release rate in $R_1$ reaches about half of that in the main CS (Fig.\,\ref{fig4}(a)).
In $R_2$, the evolutions and magnitudes of $\tilde{P}_k$ and $\tilde{P}_i$ are similar (see, dashed curves in Figs.\,\ref{fig4}(b) and (c)), which shows that magnetic energy keeps converting to kinetic and internal energies.
However, in $R_1$, the energy conversion process is relatively complicated. When a magnetic island enters $R_1$, its velocity is damped by the magnetic tension force of the flare loop.
As a result, during the second reconnection, $\tilde{P}_k$ is mostly negative (see, solid curves in Fig.\,\ref{fig4}(b)), the magnetic energy and kinetic energy significantly increase the internal energy (see positive peaks of $\tilde{P}_i$ in Fig.\,\ref{fig4} (c)).
Furthermore, the time-integrated effective changes of magnetic, kinetic, and internal energies during $t\in\left[5,15\right]$ are calculated by $\tilde{W}_m=\int_{5}^{15}\tilde{P}_m\mathrm{d}t$, $\tilde{W}_k=\int_{5}^{15}\tilde{P}_k\mathrm{d}t$, and $\tilde{W}_i=\int_{5}^{15}\tilde{P}_i\mathrm{d}t$, respectively.
For Case 4, $\tilde{W}_m$ released in $R_1$ is $13.7\%$ of that in $R_2$ (see, Tab.\,\ref{tab:Wtilde}), which shows that the total release of magnetic energy during solar flares is dominated by the reconnection in the main CS region and further replenished by that at the loop-top region.
In $R_2$, about half of released magnetic energy is converted to internal energy and half becomes kinetic energy.
However, in $R_1$, the main increment of the internal energy (about $72.5\%$) is due to magnetic energy release and the rest is from the damping of kinetic energy.

\begin{table}[h]
    \centering
    \begin{tabular}{ccc}
    \hline
    \textbf{Region} & $R_1$ & $R_2$ \tabularnewline
    \hline
    $\tilde{W}_m$ & -0.0153 & -0.1115 \tabularnewline
    $\tilde{W}_k$ & -0.0091 &  0.0527  \tabularnewline
    $\tilde{W}_i$ &  0.0211 &  0.0582  \tabularnewline
    \hline
    \end{tabular}
    \caption{Time-integrated increments of magnetic, kinetic, and internal energies in regions $R_1$ and $R_2$ for Case 4. The normalizing unit of $\tilde{W}_m$, $\tilde{W}_k$, and $\tilde{W}_i$ is $W_0=7.96\times 10^{20}\,\mathrm{erg/cm}$. Summations of $\tilde{W}_m$, $\tilde{W}_k$, and $\tilde{W}_i$ in both areas should be zero theoretically but small deviations may be caused by numerical errors.\label{tab:Wtilde}}
\end{table}

\begin{figure}
\begin{centering}
\includegraphics[scale=0.7]{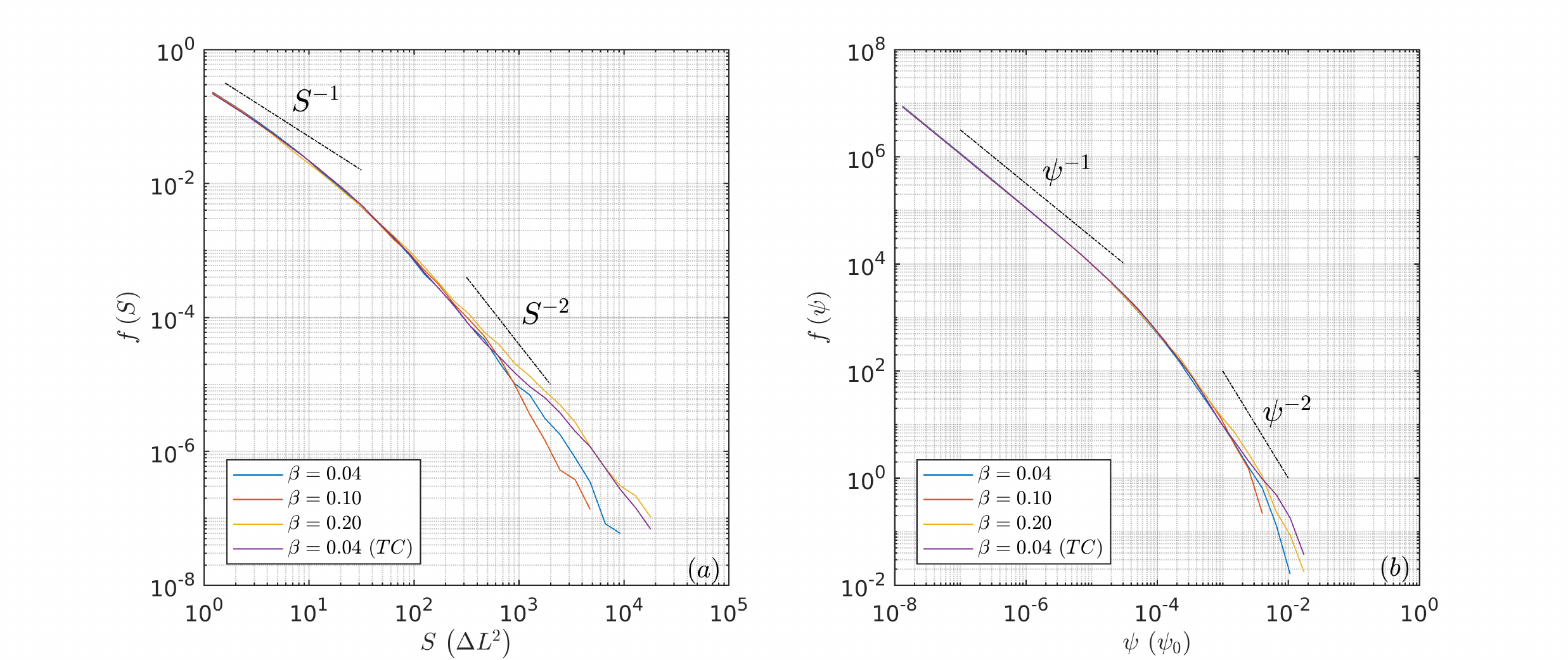}
\par\end{centering}
\caption{PDFs of magnetic island area (a) and flux (b) in the loop-top region $R_1$. To obtain the CDF of area, we set $30$ bins spaced uniformly in $\log\left(S\right)$ over the range $S\in\left[1,2.1\times 10^4\right]$ in units of $\Delta L=\Delta x\Delta y$. The configuration for calculating CDF of flux is the same as Fig.\,\ref{fig4}(b). \label{fig5}}
\end{figure}

The magnetic energy cascading at the loop-top region is closely related to the distribution of corresponding magnetic islands.
The loop-top magnetic islands generated during the second reconnection can be characterized by the evolution of cumulative distribution function (CDF) of island flux $N\left(\psi,t\right)$, which measures the number of islands with flux $\geq\psi$ at moment $t$ (Fig.\,\ref{fig4}(d)).
When the downflow magnetic islands occur the second reconnection as marked by negative peaks of $\tilde{P}_{R1}$, the CDF of island flux is significantly enhanced (see duration $t\in\left[11,11.65\right]$ in Fig.\,\ref{fig4}(d) as an example).
In Fig.\,\ref{fig5}, we analyze the probability distribution functions (PDFs) of magnetic island area and flux at the loop-top region $R_1$ based on the same method widely used for calculating plasmoid PDFs in CS \citep[e.g.,][]{ShenChengcai_2013_867,LynchBJ_2016_864,YeJing_2019_706}.
The PDF of magnetic island flux, $f\left(\psi\right)$, is determined by $f\left(\psi\right)=\mathrm{d}N_{\tau}\left(\psi\right)/\mathrm{d}\psi$, where $N_{\tau}\left(\psi\right)=\int N\left(\psi,t\right)\mathrm{d}t$ and the integration is taken over duration $t\in\left[5,15\right]$.
We also normalize the PDF to satisfy $\int f\left(\psi\right)\mathrm{d}\psi=1$.
The procedure is similar for calculating $f\left(S\right)$, the PDF of the island area.
For most of the sample domain, namely, $S<10^3$ and $\psi<10^{-3}$, both $f\left(S\right)$ and $f\left(\psi\right)$ follow a power-law independent with $\beta$ and TC.
The slopes of PDFs vary from $-1$ to $-2$ as the island scale increases.
For the large-scale end, the PDFs slightly vary for different cases, but may be unreliable because the number of large-scale islands we collected is very limited.

\subsection{Formation of termination shocks\label{subsec:ts}}

\begin{figure}
\begin{centering}
\includegraphics[scale=0.7]{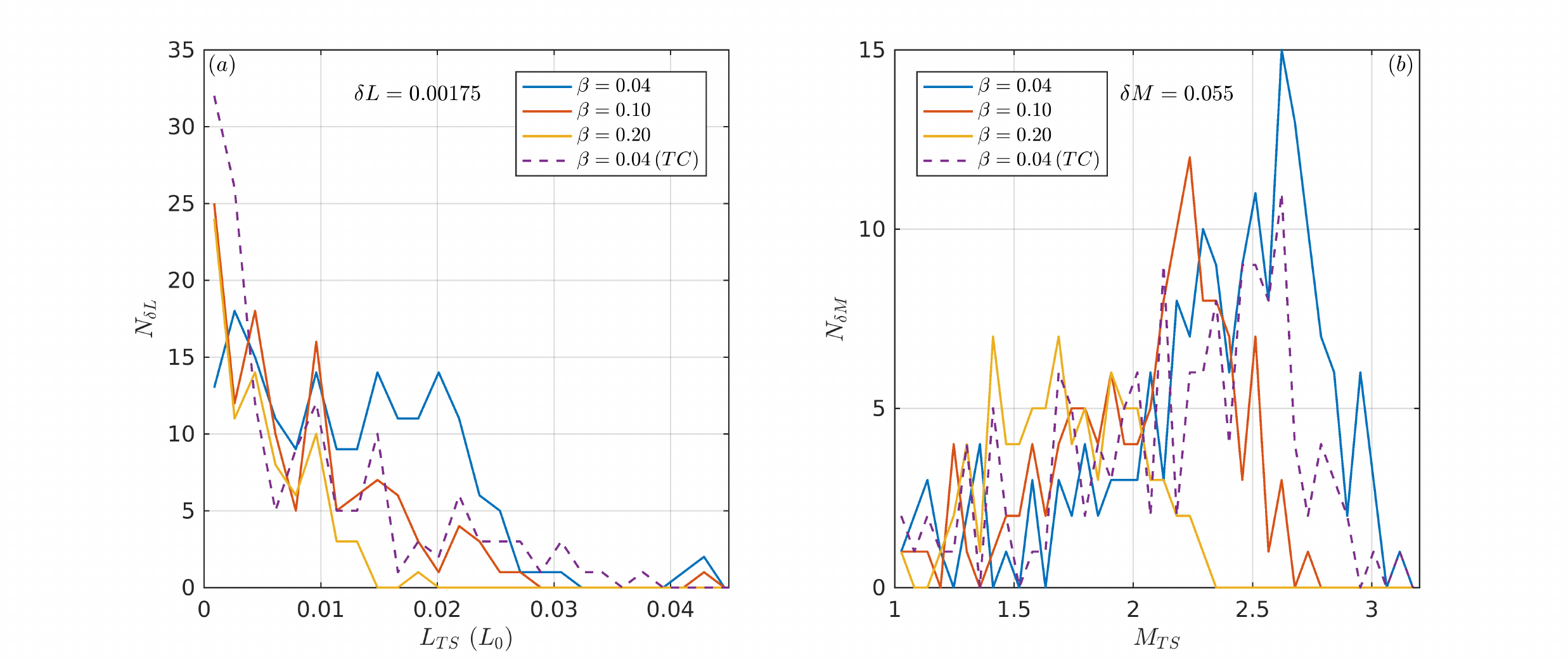}
\par\end{centering}
\caption{The histograms of spatial scale and maximum Mach number of TSs at the loop-top for all simulation cases.
As shown in Fig.\,\ref{fig3}(c) and (g), at each moment, we recognize the TS at the loop-top by a box.
Then we measure the scale along the $x$-direction, $L_{TS}$, as an approximation of the spatial scale of TS.
The strength of the TS is approximated by the maximum fast mode Mach number, $M_{TS}$, near the front of TS.
Furthermore, we uniformly sample 200 snapshots in the time period of $t\in\left[5,15\right]$ and count TS in each snapshot.
The bin width for panel (A) is $\delta L=0.00175$ and that for in Panel (b) is $\delta M=0.055$.\label{fig6}}
\end{figure}

As predicted by the standard model, the reconnection downflows may drive a TS \citep{ForbesTG_1983_783,ForbesTG_1986_785,ForbesTG_1986_786,ForbesTG_1991_787,YokoyamaTakaaki_1997_788,YokoyamaTakaaki_1998_789,YokoyamaTakaaki_2001_731,ShenChengcai_2018_705,KongXiangLiang_2019_617,KongXiangLiang_2020_719}.
Here, we further study the influences of $\beta$ and TC on the formation of TSs.
Because the TS is highly dynamic, in particular, it could be fragmented as the magnetic islands pass \citep[see,][]{ShenChengcai_2018_705}, we mainly focus on the principal TS, namely the largest and strongest segment of the TS front (see, Fig.\,\ref{fig3}(c) and (g)).
Without TC, the scale and strength of the principal TSs both decrease with the increase of $\beta$ (Fig.\,\ref{fig6}).
On the other hand, after TC is considered, although the histogram of $M_{TS}$ is similar to that without TC (see the blue and purple curves in Fig.\,\ref{fig6}(b)), the distribution of $L_{TS}$ concentrates towards the lower end.
It shows that the increase of $\beta$ and TC can suppress the scale and strength of TSs.

\section{Summary and Discussion\label{sec:sum}}

In this paper, we study the loop-top annihilation of magnetic islands formed in main vertical CS in detail.
Besides the reconnection in main eruption CS, we find that the reconnection also occurs at the flare loop top, as illuminated by \cite{BartaMiroslav_2011_822}.
The corresponding reconnection rate is found to be even of the same magnitude as that in the main CS.
Interestingly, this process is highly similar to the magnetic reconnection taking place in the earth magnetotail \citep{LuSan_2015_762,WangRongsheng_2010_761,WangRongsheng_2016_763}.
We thus suggest that the magnetic island annihilation at the flare loop top, as revealed here but neglected in the standard flare model, play a non-negligible role in further releasing magnetic energy so as to heat flare plasma and accelerate particles.

The fast magnetic reconnection in the main CS is dominated by the generation of plasmoids, which subsequently move quickly upward and downward along the CS.
This is largely in agreement with previous results \citep[e.g.,][]{BhattacharjeeA_2009_682,ShenChengCai_2011_619,ShenChengcai_2018_705,KongXiangLiang_2019_617,KongXiangLiang_2020_719} and implies the existence of fine structures in the CS during solar eruptions.
The second reconnection characterized by horizontal CSs is initiated when magnetic islands collide with the flux above the flare loops.
As the magnetic islands moving towards the flare loop, the reconnection rate first grows and then decreases with the peak being comparable with that in the main CS.
The second reconnection also generates smaller-scale secondary magnetic islands which enhance the turbulence and thus further facilitate the release of magnetic energy at the flare loop top.
Moreover, the whole reconnection process seems to be intermittent because of the presence of magnetic islands, it is thus expected that the acceleration of non-thermal particles is intermittent. This might be used to explain the QPPs observed during flares.

The cascading law of magnetic islands at the loop top seems to be less affected by background $\beta$ or TC.
The PDF of the magnetic island flux generated during the second reconnection follows a power-law varying from $f\left(\psi\right)\sim\psi^{-1}$ to $\psi^{-2}$, similar to the PDF of the magnetic island flux in the main CS obtained by \cite{LoureiroNF_2012_856}, \cite{ShenChengcai_2013_867} and \cite{LynchBJ_2016_864}.
However, for the spatial scale and strength of the TSs formed by the downward outflows or plasmoids, it is found that both of them decrease with the increase of plasma beta or including TC.

It is worthy of mentioning that simulations of magnetic reconnection depend on the spatial resolution. The results presented here are obtained under a high-resolution spatial grid ($3840\times 7680$), which can realize the simulations with background $S=10^6$ as discussed by \cite{YeJing_2020_653}.
The thickness scales of CSs in both $R_1$ and $R_2$ are on the order of $0.01$ which is $0.5\%$ of the $x$-domain (see, Fig.\,\ref{fig1} and Fig.\,\ref{fig3}).
Under the current resolution, the CS thickness can be resolved by about 20 grids.
Furthermore, the scale of internal singular layer in classic tearing mode theory is $\delta\sim aS_a^{-1/4}$ \citep{PucciFulvia_2017_894}, where $a$ denotes the width of a current sheet, $S_a=av_A/\eta$ is the locale Lundquist, and $v_A$ is the Alfv\'{e}n speed.
The number of grids resolving the inner singular layer can be estimated by
\begin{equation}
    \frac{\delta}{\Delta L}\sim\frac{aS_a^{-1/4}}{\Delta L}=\frac{\eta^{1/4}}{\Delta L}\frac{a^{3/4}}{v_A^{1/4}}\,,
\end{equation}
where $\Delta L$ denotes the dimensionless size of mesh grid.
Setting $\Delta L=2/3840$, $\eta=5\times10^{-6}$, $v_A=1$, and $a=0.01$, we have $\delta/\Delta L\sim 3$.
It's shown that the small-scale inner singular layer can still be roughly resolved.
We have also studied the influence of resolutions on our conclusions by running cases with different grid configurations. It is shown that the valid conclusions are only for relatively high-resolution simulations.

In our numerical model, the fine dynamics of the chromosphere might be unreal, because the effects of partially ionized plasma (PIP) is not included.
It is true that partial ionization has influences on the properties of shocks, the CS structure, and the fractal reconnection \citep{ImadaS_2011_886,HillierA_2016_885}.
However, in this paper, we mainly focus on the reconnection dynamics in the high-temperature corona.
Thus, the effects of partially ionized chromosphere on our results are very limited.
The only influence is on the structure and dynamics in the post-flare loops where the chromospheric plasma is evaporated into.

\acknowledgments

We would like to thank the anonymous referee for valuable suggestions. This research is supported by Natural Science
Foundation of China grants 11722325, 11733003, 11790303, 11790300, and 11805203.

\bibliography{refs}{}
\bibliographystyle{aasjournal}

\end{document}